\documentclass[12pt]{article}
\usepackage{epsfig}
\usepackage{pstricks}

\begin{document}

\title{Theory of self-diffusion in GaAs}

\author{  Michel~Bockstedte and Matthias~Scheffler\\
Fritz-Haber-Institut, Faradayweg 4-6, D-14195 Berlin-Dahlem, Germany}
\date{October 4, 1996}


\maketitle
\begin{abstract}
  {\em Ab initio} molecular dynamics simulations are employed to investigate
  the do\-mi\-nant migration mechanism of the gallium vacancy in GaAs as well as
  to assess its free energy of formation and the rate constant of gallium
  self-diffusion. Our analysis suggests that the vacancy migrates by second
  nearest neighbour hops. The calculated self-diffusion constant is in good
  agreement with the experimental value obtained in $^{69}{\rm GaAs/}^{71}{\rm
    GaAs}$ isotope heterostructures and at significant variance with that
  obtained earlier from interdiffusion experiments in
  GaAlAs/GaAs-heterostructures.\\[1cm]
  \begin{tabbing}
  {\bf KEYWORDS}:\ \=III-V compounds, self-diffusion, vacancy
  formation, vacancy mi-\\\>gration, {\em ab initio} molecular dynamics
  \end{tabbing}
\end{abstract}

\section{Introduction}
Gallium self-diffusion in GaAs is a fundamental process with important
technological implications as to e.g. degradation of heterostructure devices.
It is experimentally well established that this diffusion process is mediated
by gallium vacancies~\cite{tan:91,deppe:88}. The gallium self-diffusion
constant is hence proportional to the diffusivity of gallium atoms mediated by
the migration of a single vacancy and to the concentration of gallium
vacancies. However, the mechanism of vacancy migration is still unknown.
Migration of a gallium vacancy may involve atoms on either of the two
sublattices, possibly transforming one point defect into complexes of others.
In fact, it has not been identified so far whether the diffusion involves
only the gallium sublattice or whether it proceeds by successive nearest
neighbour hops between the two sublattices.

The gallium self-diffusion constant deduced from interdiffusion experiments in
mul\-tiple quantum well structures is generally associated with an activation
energy of $6\,{\rm eV}$~\cite{tan:91,rouviere:92}.  Rouviere et
al.~\cite{rouviere:92} extracted an unusually high value for the vacancy
formation entropy of $32.9\,k_{\rm B}$ from interdiffusion data compatible with
a diffusion entropy of $25\,k_{\rm B}$ obtained earlier~\cite{tan:91}.  This
is in contrast to a more recent study for GaAs isotope heterostructures. For
these it was found~\cite{wang:96} that gallium self-diffusion has an
activation energy of $\sim 4\,{\rm eV}$ and an entropy of diffusion $\sim 8
k_{\rm B}$.

In the present {\em ab initio} molecular dynamics study we investigate the
microscopic aspects as well as the thermodynamical properties of the gallium
vacancy diffusion.  In section 3 we calculate the free energy of formation and
the formation entropy of the gallium vacancy, which determine its
concentration at thermal equilibrium.  In section 4 we investigate the question for the dominant
migration mechanism by {\em ab initio} molecular dynamics simulations of the
motion of the neutral vacancy.  Guided by our simulations we discuss the
diffusion events identified therein also including the effects of the Fermi
level.  In section 5, the rate constant for the dominant mechanism is finally
evaluated employing {\em ab initio} molecular dynamic simulations and the self-diffusion
constant is calculated.
\section{Computational Method}
Diffusion in covalently bonded materials implies breaking and making of bonds
of the migrating atom to its neighbours. This requires a method to
quantitatively describe the bonding and re-bonding process along all possible
migration paths. Density functional theory~\cite{hohenberg:64} together with a
reliable approximation to exchange and correlation provides a sound basis
for such a study. We employ the local density approximation to the
exchange-correlation functional~\cite{ceperley:80,perdew:81}. The electron
ground state is obtained by solving the Kohn-Sham equation~\cite{kohn:65}, in
which the core electrons are treated by the frozen-core approximation and the
corresponding ion cores are replaced by fully-separable~\cite{kleinman:82}
norm-conserving pseudopotentials~\cite{hamann:89}. The defect system is
represented by a 63 atom super cell, which is periodically repeated on a
simple cubic lattice.  The Kohn-Sham orbitals are expanded in a plane wave
basis set including plane waves up to an energy cut-off of 8\,Ry.

To study the motion of the gallium vacancy and to calculate ensemble averages
related to the entropy of formation and the rate constant we perform detailed
{\em ab initio} molecular dynamics simulations on the Born-Oppenheimer
surface~\cite{bockstedte:96a}. In this approach the dynamics of the gallium
and arsenic nuclei is treated by classical mechanics, i.e. the motion of the
nuclei is governed by Newton's equation of motion. For any configuration of the
nuclei we first find the electron ground state using density functional theory
and calculate the forces. We then integrate the
equation of motion of the nuclei for the next time step using the Verlet
algorithm~\cite{verlet:67}.  This allows for a time step of
$\frac{1}{12}\,\nu_{\Gamma{\rm O}}^{-1}$ -- where $\nu_{\Gamma{\rm
    O}}=8.8\,{\rm THz}$ is the highest phonon-frequency in GaAs. This time
step is much longer (by a factor 27~\cite{zhang:90}) than that used in the
Car-Parrinello method ~\cite{car:85}. The simulations are carried out at
constant volume. A Nos\'e-Hoover thermostat~\cite{nose:84,hoover:85} enables
the inclusion of temperature allowing for the calculation of canonical
ensemble averages.

In order to analyse the observed diffusion events and to assess migration
barriers we have calculated adiabatic potential energy surfaces. In these
calculations the adiabatic potential energy surface is obtained as a function
of the relevant migration coordinates by constraining these coordinates
and allowing for the relaxation of the other degrees of freedoms.

\section{Free energy of formation}

In thermodynamic equilibrium and at constant volume the concentration of
intrinsic point defect is determined by the free energy of formation $F_{\rm
  f}$
\[
c_{\rm D}=c_{\rm S}\,{\rm exp}\left({-\frac{F_{\rm f}}{k_{\rm
B}\/T}}\right)\quad,
\] where $c_{\rm S}$ is the concentration of sites in the crystal open to the
defect~\cite{scherz:93}.  GaAs is a compound material and in thermal
equilibrium the stoichiometry of the crystal is determined by the chemical
environment. For example, an arsenic overpressure results in a higher
concentration of arsenic antisite defects and gallium vacancies. The chemical
potential of arsenic and of gallium may vary only within certain bounds in
order to preserver the stability of the GaAs crystal. Otherwise the crystal
would decompose into more stable gallium and arsenic phases. The free energy
of defect formation is given by
\begin{equation}
F_{\rm f}=F_{\rm D,cell}-n_{\rm e}\,\mu_{\rm F}-n_{\rm Ga}\,\mu_{\rm Ga}-n_{\rm
  As}\,\mu_{\rm As}\quad,
\end{equation} where $F_{\rm D,cell}$ is the free energy of the super cell
containing the defect, $n_{\rm As}$ arsenic atoms and $n_{\rm Ga}$ gallium atoms;
$\mu_{\rm As}$ and $\mu_{\rm Ga}$ are the chemical
potentials of the two species. Depending on the charge state of the defect $n_{\rm e}$ electrons
are exchanged between the electron reservoir  (the Fermi level) $\mu_{\rm F}$
and the defect levels. In thermal equilibrium the chemical potentials of the
gallium reservoir, the arsenic reservoir and the bulk phase of GaAs are
related: $\mu_{\rm GaAs}=\mu_{\rm Ga}+\mu_{\rm As}$. Therefore we consider
only the deviation of the arsenic chemical potential from the value of an
arsenic-crystal: $\Delta\mu=\mu_{\rm As}-\mu_{\rm As,bulk}$. With this
definition the free energy is expressed as
\begin{equation}
  F_{\rm f}=F_{\rm D}-n_{\rm e}\,\mu_{\rm F}-(n_{\rm As}-n_{\rm
    Ga})\Delta\mu\quad,
\end{equation} where $F_{\rm D}=F_{\rm D,cell}-n_{\rm Ga}\mu_{\rm
  GaAs}-(n_{\rm As}-n_{\rm Ga})\mu_{\rm As,bulk}$ and $F_{\rm D}-n_{\rm
  e}\,\mu_{\rm F}$ is the free energy of formation with respect to a reservoir
of arsenic bulk material. The allowed range for $\Delta\mu$ is estimated
considering the bulk phases of gallium and arsenic as the limiting
cases~\cite{qian:88}. This corresponds to the range $-\Delta H\leq\Delta\mu
\leq 0$, where $\Delta H$ is the heat of formation of GaAs.

Thermodynamic integration together with {\em ab initio} molecular dynamics
simulations provides a method to accurately calculate the free energy
differences. It accounts for the vibrational contribution to the free
energy differences also including anharmonic contributions. In this method
the coupling of a gallium atom to the other atoms in the simulation cell is
reduced by means of a coupling parameter $\lambda$ in order to create a
vacancy. The value $\lambda=0$ refers to the perfect crystal. At the same
time this atom is coupled to a harmonic oscillator to confine its motion in
the cell. The decoupling is accomplished by reducing the strength of the
pseudopotential of this atom by the factor $(1-\lambda)$, $0\leq\lambda\leq
1$ and by removing the corresponding fraction of electrons from the highest
occupied defect levels. The free energy difference $\Delta F=F(1)-F(0)$ is
then obtained by
\begin{equation}
\Delta F=\int^{1}_{0}\,\frac{\partial\/F}{\partial\/\lambda}\,d\/\lambda,
\label{free_energy}
\end{equation} where
$\partial\/F/\partial\/\lambda=\left\langle\partial\/E_{\rm tot}/
\partial\/\lambda\right\rangle$ and $E_{\rm tot}$ is the total energy of 
the system. The derivative of the total energy with respect to the coupling
parameter $\partial\/E_{tot}/\partial\/\lambda$ is readily evaluated and the
canonical averages are calculated by molecular dynamics simulations. The
numerical integration in Eq. (\ref{free_energy}) is performed by Gaussian
integration and requires the evaluation of
$\left\langle\partial\/E_{tot}/\partial\/\lambda\right\rangle$
typically at four values of $\lambda$.  We have calculated $\Delta F$ for the
neutral vacancy at $600^{\circ}{\rm C}$.  The averages are calculated for
simulations as long as $12\,{\rm ps}$ and the overall statistical uncertainty
in $\Delta F$ is $\sim 1\,k_{\rm B}\/T$.  Having calculated $\Delta F$, $F_{\rm
  D}$ is readily obtained from
\begin{equation}
  F_{\rm D}=\Delta F-F^{\rm osc}_{\rm Ga}+\mu_{\rm GaAs}-\mu_{\rm As,
    bulk}\quad,
\end{equation} where $F^{\rm osc}_{\rm Ga}$ is the free energy of the neutral
gallium atom coupled to the oscillator potential. The chemical potentials of
GaAs bulk and As bulk are obtained from total energy calculations together
with the corresponding entropies, which we have calculated from the phonon
spectrum in the case of GaAs bulk and taken from calorimetric
measurements~\cite{anderson:30} in the case of the As bulk. We thus obtain
$F_{\rm D}=2.8\,{\rm eV}-7.3\,k_{\rm B}\/T$. The resulting value for the
entropy of the neutral vacancy is $S_{\rm D}=7.3\,{k_{\rm B}}$, which is
within the range $5\,k_B \leq\,S_{\rm f}\,\leq\,9\,k_B$, estimated by Bernholc
{\em et~al.}~\cite{bernholc:95} for the triply negative vacancy
by a similar approach.


In the interdiffusion experiments by Rouviere {\em et~al.}~\cite{rouviere:92}
in $n$-type GaAlAs/GaAs multi layer quantum well structures ${\rm As}_4$-gas
was considered as the relevant reservoir. The gallium vacancy (${\rm V}_{\rm
Ga}$) formation is then governed by the reaction
\begin{equation}
\frac{1}{4}\,{\rm
    As}_{4\,{\rm(gas)}}\rightleftharpoons \,{\rm V}_{\rm Ga}+{\rm GaAs}_{\rm(bulk)}\quad.
\label{react}
\end{equation} The entropy of formation deduced from these
experiments amounts to $S^{\rm exp}_{\rm f}=32.9\,k_{\rm B}$. The ideal gas
contribution of the ${\rm As}_4$-gas to the entropy has been explicitly
removed from the value and for comparison we do not include it as well. In
$n$-type material the vacancies are created in a triply negative charge state
and we approximate the vibrational contribution to the formation entropy by
that of the neutral vacancy.  For the reaction in Eq. (\ref{react}) we thus obtain an
entropy of $S_{\rm f}=8\,k_{\rm B}$. The experimental value $S^{\rm exp}_{\rm
  f}$ is four times larger than our value.  Note, that there is a good
agreement between experiment and theory in the case of the silicon vacancy and
that the formation entropies of the vacancy in silicon and the gallium vacancy
in GaAs are similar. The experimental value has been determined from
interdiffusion data in GaAlAs/GaAs multi layer quantum wells together with
three other parameters describing the interdiffusion constant by a
simultaneous fit. In the model it was assumed that the interdiffusion is
essentially described by gallium self-diffusion.  In fact the high
experimental value may be due to other processes than the formation of gallium
vacancies or even due to difficulties in the experimental analysis related to
the small temperature range between $600^{\circ}{\rm C}$ and $650^{\circ}{\rm
  C}$ in which experimental data have been obtained.

\section{The gallium vacancy migration path}
Migration of atoms in solids is a fairly complex process. Especially in
compound semiconductors like GaAs hops between the gallium and arsenic
sublattice always imply the formation of antisite defects. Migration of the
gallium vacancy may solely proceed on the gallium sublattice or by nearest
neighbour hops as has been proposed by Van Vechten~\cite{vanvechten:84}.  A
migration mechanism involving nearest neighbour hops consists of a few
intermediate stages, where in the final stage all the antisite defects are
removed. Otherwise such a mechanism would create an unfavourable excess of
antisite defects beyond the equilibrium concentration. 

In order to learn about the likely events in vacancy diffusion we have
performed molecular dynamics simulations of the neutral vacancy close to the
melting temperature of the crystal and at constant volume. The volume of the
simulation cell was fixed at the theoretical lattice constant.  Our
simulations identify several interesting phenomena. For example, we find that
nearest neighbour hops (i.e., from the As sublattice into the Ga vacancy)
occur with a rather high probability.

In Fig.~\ref{nn_hop} the trajectories of the atoms in the GaAs zig-zag chain
containing the gallium vacancy and the hopping arsenic atom are plotted as a
projection onto the (110)-plane. The simulation has been performed at a
temperature of $1600\,{\rm K}$ and the part shown corresponds to a
simulation time of $8\,{\rm ps}$. The figure shows that the arsenic atom has
jumped into the vacancy.  Thereby it leaves an arsenic vacancy
(${\rm V}_{\rm As}$) behind and an arsenic antisite-vacancy complex (${\rm
As}_{\rm Ga}$-${\rm V}_{\rm As}$) is created.  Yet, after a short time
($\sim 2\,{\rm ps}$) the arsenic atom returns to its initial site. 
For the short time when the ${\rm As}_{\rm Ga}$-${\rm
V}_{\rm As}$ complex exists we find attempts of the three nearest neighbour
gallium atoms to hop into the vacancy.  However, analysing the potential
energy surface as a function of the displacement of one of the gallium
neighbours 
we find a plateau at the typical bond distance instead of a local minimum.
Hence such a hop is impossible in the neutral charge state.

In a simulation at a temperature of $1700\,{\rm K}$ we observe an attempt of a
second nearest neighbour hop.  The gallium atom leaves its lattice site and
approaches the ideal interstitial site as close as $1.5\,{\rm bohr}$. An
analysis of the time-evolution of the bond distances to the four nearest
neighbours reveals that the bond distance of three of these atoms remains
effectively unchanged during this event. Only the distance to the neighbouring
arsenic atom farthest away from the vacancy increases by $3\,{\rm bohr}$ and
eventually becomes as large as $7\,{\rm bohr}$.

We proceed by further analysing the events identified in the simulations.  The
transition state and the migration energy barrier associated with these events
are evaluated by calculating the adiabatic potential energy surfaces for the
coordinates we have identified supported by the simulations. In a reaction of
intrinsic defects the initial defect and the final defect may be in a different
charge state for a given position of the Fermi level~\cite{baraff:86}. We also
include this effect in our discussion and evaluate the adiabatic potential
energy surfaces for different charge states of the defect system. The migration
energy barrier then depends on the position of the Fermi level. 

First we discuss the nearest neighbour hop as function of the Fermi
level. The coordinate $\xi_{\rm nn}$
\begin{equation}
  \xi_{\rm nn}=\left({\bf R}_{\rm As}-\frac{1}{3}\sum_{\rm nn}\,{\bf R}_{{\rm
      Ga}_{i}}\right) \cdot {\bf e}_{(111)}
\end{equation} 
describes the position of the arsenic atom on the ideal bond
axis with respect to the three neighbouring gallium atoms, where ${\bf
  e}_{(111)}$ is the unit vector in the (111)-direction. This coordinate is
compatible with what we have inferred from our simulations and it fulfils the
conditions~\cite{bennett:77} for a migration coordinate. We have calculated
the potential as a function of this coordinate for the relevant charge states
of the defect minimising the total energy with respect to all other
coordinates
\begin{equation}
  E(\xi_{\rm nn},n_{\rm e})=\min_{\xi_{\rm nn}={\rm const.}}E\left(\{{\bf
    R}_i\}\right)-n_{\rm e}\,\mu_{F}\quad,
\end{equation} where $n_{\rm e}$ is the number of additional electrons in the
defect levels and $\mu_{\rm F}$ is the Fermi level. Note that only one
coordinate is constrained and the resulting migration path of the arsenic atom
not necessarily coincides with the bond axis.

The resulting energy surfaces corresponding to negative charge states of the
defect have only one local minimum at the configuration where the vacancy is
at the gallium site.  The $({\rm As}_{\rm Ga}{\rm -}{\rm V}_{\rm As})$-complex
is thus instable in a negative charge state. For the neutral and positive
charge states there exist two local minima and the $({\rm As}_{\rm Ga}{\rm
  -}{\rm V}_{\rm As})$-complex is stable in these charge states.

In Fig.~\ref{nn_hop_fermi} the function $E(\xi_{\rm nn},n_{\rm e})$ is
plotted for two different positions of the Fermi level: close to the
conduction band minimum and at mid gap. For a Fermi level close to the
conduction band minimum only the negatively charged defect is favourable and
as discussed above the $({\rm As}_{\rm Ga}{\rm -}{\rm V}_{\rm As})$-complex
is instable. Hence a nearest neighbour hop is impossible. With a Fermi level
at mid gap the gallium vacancy is in a triply negative charge state whereas
for the $({\rm As}_{\rm Ga}{\rm -}{\rm V}_{\rm As})$-complex the local
minimum corresponds to the neutral charge state.  This energy is $1.8\,{\rm eV}$
higher than that of the gallium vacancy. Therefore, during the hop of the
arsenic atom into the gallium vacancy three electrons need to be transferred
to other distant defect states. The
minimum barrier for the hop is $2.1\,{\rm eV}$, assuming the three excess
electrons are transferred instantaneously to the reservoir.
For a Fermi level close to the valence band maximum the $({\rm As}_{\rm
  Ga}-{\rm V}_{\rm As})$-complex is in a triply positive charge state, i.e.
all electrons are removed from the localised defect levels in the band gap.
The gallium vacancy is under this condition by $0.6\,{\rm eV}$ energetically
less favourable than the complex.

Most experiments for gallium self-diffusion~\cite{tan:91,deppe:88,wang:96}
have been performed in $n$-type or intrinsic material. This hop is
impossible in $n$-type material.
Hence, we do not consider this mechanism as a relevant
diffusion mechanism.

Guided by our simulations we develop in the following a microscopic picture of
the second nearest neighbour hop. We have seen that the gallium atom in moving
into the interstitial region breaks its bond to the nearest neighbour arsenic
atom farthest away from the vacancy. In passing on it leaves its lattice site
vacant and hops half way between the two vacant lattice sites through a
plane perpendicular to the (110)-direction.  This plane is a plane of mirror
symmetry of the problem, i.e. if a migration path of the system through a
saddle point is not symmetric with respect to this symmetry than there exist
two paths related to each other by symmetry. The four gallium atoms indicated
in Fig.~\ref{2nn_hop}a are located in this plane and form a gate through which
the hopping atom has to pass. We take the centre of this gate as the origin
and consider the following coordinate for the gallium atom
\begin{equation}
\xi_{\rm Ga}=\left({\bf R}_{\rm Ga}-
\frac{1}{4}\sum_{\rm gate}{\bf R}_{{\rm Ga}_i}\right)
\cdot {\bf e}_{(110)}\quad,
\label{coor_ga}
\end{equation} where ${\bf e}_{(110)}$ is the unit vector in the
(110)-direction. Due to the inward relaxation of the arsenic neighbours of the
vacancy, the arsenic atom to which the hopping gallium atom is bonded sticks
out of the plane by $0.4\,{\rm bohr}$. During the hop this atom has to pass
onto the opposite side of the gate. However, the correlation of this motion
with that of the gallium atom and the path are {\em a priori} not
known. Therefore we also include the coordinate $\xi_{\rm As}$ defined similar
to $\xi_{\rm Ga}$ in Eq. (\ref{coor_ga}) in the calculation of the potential
energy surface.  The potential energy surface is obtained by constraining the
two coordinates and minimising the total energy with respect to all other
degrees of freedom of the neutral system
\begin{equation}
E(\xi_{\rm Ga},\xi_{\rm As})=
\min_{\xi_{\rm Ga},\xi_{\rm As}={\rm const.}}\,E_{tot}\left(\{{\bf R}_i\}\right).
\end{equation} The result is shown in Fig.~\ref{2nn_hop}b. The stars indicate
the minima corresponding to the vacancy being on either side of the gate. We
find two saddle points marked by crosses. These saddle points constitute the
bottle neck through which the system has to pass in hopping to the other
minimum.  At the saddle points the gallium atom is located in the gate while
the arsenic atom moves by $1.5\,{\rm bohr}$ out of the plane to either
side. The migration barrier for the neutral system is $1.7\,{\rm eV}$. The
configuration with the gallium atom and the arsenic atom located in the gate
at the same time is by $0.3\,{\rm eV}$ higher in energy. It is the local
maximum on the potential energy surface located between the two saddle
points.

There are two migration paths connecting the two minima. Along one path the
gallium atom pushes the arsenic atom out of the gate in the direction of the
hop. When the gallium atom has passed through the gate both atoms are
located on the same side of the plane till the arsenic atom finally moves to
the other side.  Along the other path the gallium atom first pulls the
arsenic atom through the gate before passing it.  By time reversal symmetry
and by mirror symmetry with respect to the plane these two paths are in fact
equivalent. This corresponds to the point symmetry with respect to the
origin of the potential energy surface in Fig~\ref{2nn_hop}b.  However, if
the temperature is sufficiently high both atoms may pass with a high
probability through the gate at the same time. The bottle neck then also
includes the local maximum at $(\xi_{\rm Ga}=0,\xi_{\rm As}=0)$ and the
dividing surface separating the equivalent minima may be described to a good
approximation by the plane ($\xi_{\rm Ga}=0$).

We have also calculated the migration barrier for the triply negative charge
state. The configuration at the saddle points corresponds to that of the
neutral system. The migration barrier in this charge state is $1.9\,{\rm eV}$
compared to $1.7\,{\rm eV}$ for the neutral system. Hence for the second
nearest neighbour hop Fermi level effects are negligible.

\section{Rate constant of the second nearest neighbour hop}

As discussed above we consider the second nearest neighbour hop as the
relevant mechanism of vacancy migration. Therefore we calculate the rate
constant for this mechanism.

Since the migration barrier is fairly high the occurrence of such a hop is a
rare event, even at high temperatures. The rate constant of such a rare event
is well described by transition state theory~\cite{vineyard:57}. Most of the
neighbours of the vacancy will oscillate at their lattice site and the energy
of the system will be close to the local minimum. Only seldom a second nearest
neighbour hops through the dividing surface into the vacancy. The reduction of
this surface for the two relevant coordinates is indicated by the dashed line
in the contour plot of Fig.~\ref{2nn_hop}b. However, this dividing surface is
high dimensional and involves also the other 187 degrees of freedom of the
super cell, which are contributing to the entropy of the event.

At low temperatures the migrating system will closely follow the migration
path depicted in Fig.~\ref{2nn_hop}b. The two dividing points are equivalent
by symmetry and the overall rate constant is twice as large as the rate
constant for one of the two migration paths. At high temperatures paths
through the dividing surface passing in the vicinity of the local maximum have
a considerably higher weight than at lower temperatures. In this case the two
migration paths loose their importance and the migration is described by a
single coordinate.  As discussed in the previous section we consider the
coordinate $\xi_{\rm Ga}$ as the migration coordinate in the calculation of
the rate constant.

Since paths through the dividing surface deviating considerably from the two
migration paths have relevance as discussed above, we also include the
anharmonicity of the potential energy surface in the calculation of the rate
constant. The method suggested by Paci {\em et~al.}~\cite{paci:92} enables such
calculation. In this method the corresponding ensemble averages are evaluated
by molecular dynamics simulations considering all coordinates of the system.
We have calculated the rate constant for the neutral vacancy at
$600^{\circ}{\rm C}$ by this method.  The value we obtain is
\begin{equation}
\Gamma_{\rm TST}=6705\,s^{-1}\quad.
\end{equation} With an energy barrier $E_{\rm m}=1.7\,{\rm eV}$
this corresponds to $\Gamma_{\rm TST}= 46\,{\rm
  exp}\left({-{1.7\,{\rm eV}}/{k_{\rm B}\/T}}\right)\,{\rm THz}$. The prefactor
  $\Gamma_0=46\,{\rm THz}$ is a factor $5.5$ larger than the highest
  phonon frequency in GaAs by which this prefactor is usually approximated in
  lack of a better value.

The gallium self-diffusion constant is obtained from the free energy of
formation $F_{\rm f}$ and the rate constant
$\Gamma_{TST}$ by
\begin{equation}
  D_{\rm Ga}= \frac{1}{6}\, n\,d^2\,\Gamma_{\rm TST}\,\,{\rm
    exp}\left({-\frac{F_{f}}{k_{\rm B}\/T}}\right),
\label{diffusion}
\end{equation} where $n$ is the number of second nearest neighbour sites
($n=12$) and $d$ is the distance between these sites ($d=a/\sqrt{2}$, where
$a$ is the lattice constant). Gallium self-diffusion has been mainly
investigated in material grown in an arsenic-rich environment and for $n$-type
doping~\cite{tan:91,deppe:88}. Thus the triply negative vacancy ${\rm V}_{\rm
  Ga}^{3-}$ has to be considered in Eq. (\ref{diffusion}).  The free energy of
formation in arsenic rich environment $F_{\rm f}=3.6\,{\rm eV}-7.3\,k_{\rm
  B}\/T-3\,\mu_{\rm F}$ depends on the position of the Fermi level and
consequently also the diffusion constant.  The experimental self-diffusion
constant is therefore related to intrinsic conditions as a reference. Under
these conditions the activation energy, given by the sum of the formation
energy $E_{\rm f}=2.1\,{\rm eV}$ and the migration barrier $E_{\rm
  m}=1.9\,{\rm eV}$, amounts to $4\,{\rm eV}$. For the activation energy Chen
{\em et~al.}~\cite{chen:94} and Dabrowski {\em et~al.}~\cite{dabrowski:94a}
arrived at the same value of $4\,{\rm eV}$. The prefactor $D_0$, which is
determined by $D_0=2\/d^2\,\Gamma_0\,{\rm exp}\left({S_{\rm f}}/{k_{\rm
    B}\/T}\right)$ amounts to $D_0=208\,{\rm cm}^2\/{\rm s}^{-1}$,
approximating $\Gamma_0$ by the value obtained for the neutral system. The
gallium self-diffusion constant we obtain is $D_{\rm Ga}=208\,{\rm
  exp}\left(-{4\,{\rm eV}}/{k_{\rm B}\/T}\right)\,{\rm cm}^2\/{\rm s}^{-1}$.
In the Arhenius-plot of the self-diffusion constant of Fig.~\ref{diff} we
compare our result with the fit by Tan {\em et~al.}~\cite{tan:88} to
experimental interdiffusion data from GaAlAs/GaAs-heterostructures and with
the fit to self-diffusion data obtained in $^{69}$GaAs/$^{71}$GaAs
isotope-heterostructures by Wang {\em et~al.}~\cite{wang:96}. The result for
the interdiffusion constant by Tan {\em et~al.}~\cite{tan:88} is described by
an activation energy of $6\,{\rm eV}$ and a prefactor of $2.9\times
10^{8}\,{\rm cm}^2\/{\rm s}^{-1}$.  It clearly deviates from our theoretical
result and the experimental result obtained in isotope heterostructures by
Wang {\em et~al.}~\cite{wang:96}. Whereas the latter, described by an
activation energy of $4.2\,{\rm eV}$ and a prefactor of $43\,{\rm cm}^2\/{\rm
  s}^{-1}$, is in good agreement with our result.
\section{Summary and conclusion}
The gallium self-diffusion in GaAs as mediated by the gallium vacancy has been
investigated by means of {\em ab initio} molecular dynamics simulations. We
have employed this technique as a microscope to analyse the motion of the
vacancy and to assess the free energy of vacancy formation and the
self-diffusion constant.  It has been shown that the vacancy migrates by
second nearest neighbour hops solely on the gallium sublattice. The pronounced
Fermi level dependence of the migration barrier of the nearest neighbour hop
lets us exclude a migration mechanism by nearest neighbour hops for the
relevant experimental conditions. For a Fermi level close to the conduction
band minimum the hop is even impossible.  The gallium self-diffusion constant
has been obtained from the calculated free energy of formation and the rate
constant of the second nearest neighbour hop. Our result disagrees with a fit
by Tan {\em et~al.} to interdiffusion data obtained in GaAlAs/GaAs
heterostructures.  However, it is in good agreement with self-diffusion data
measured by Wang
{\em et~al.} in isotope heterostructures. 

\clearpage
\noindent {\bf Figure captions}\\[1.5cm] 

\begin{figure}
\epsfig{file=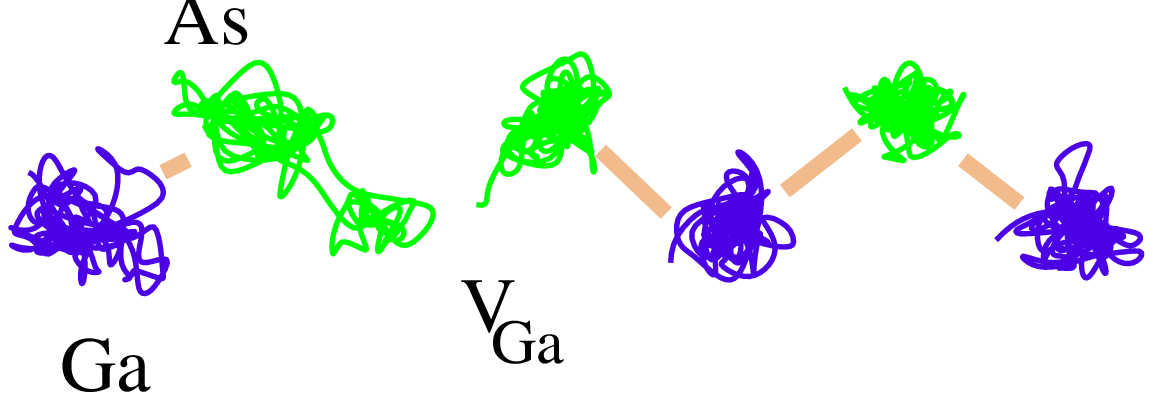,angle=-90,width=14cm}
\vspace*{4cm}
\caption{\hspace*{15cm}}
\label{nn_hop}
\clearpage

\end{figure}

Fig.~\ref{nn_hop}:\\ Nearest neighbour hop of an arsenic atom into the vacancy.
Trajectories are projected onto the (110)-plane containing the gallium vacancy and the
hopping arsenic atom.\\[1cm] 

Fig.~\ref{nn_hop_fermi}:\\ Nearest neighbour hop: Potential energy vs.
reaction coordinate for the relevant charge states for a Fermi level
corresponding to $n$-type conditions and a Fermi level at mid gap.\\[1cm]

\begin{figure}
  \pspicture(20,8.3)
  \rput{0}(0.0,2.5){\large (a)}
  \rput{0}(7.5,2.5){\large (b)}
  \rput{0}(3.0,5){\epsfig{figure=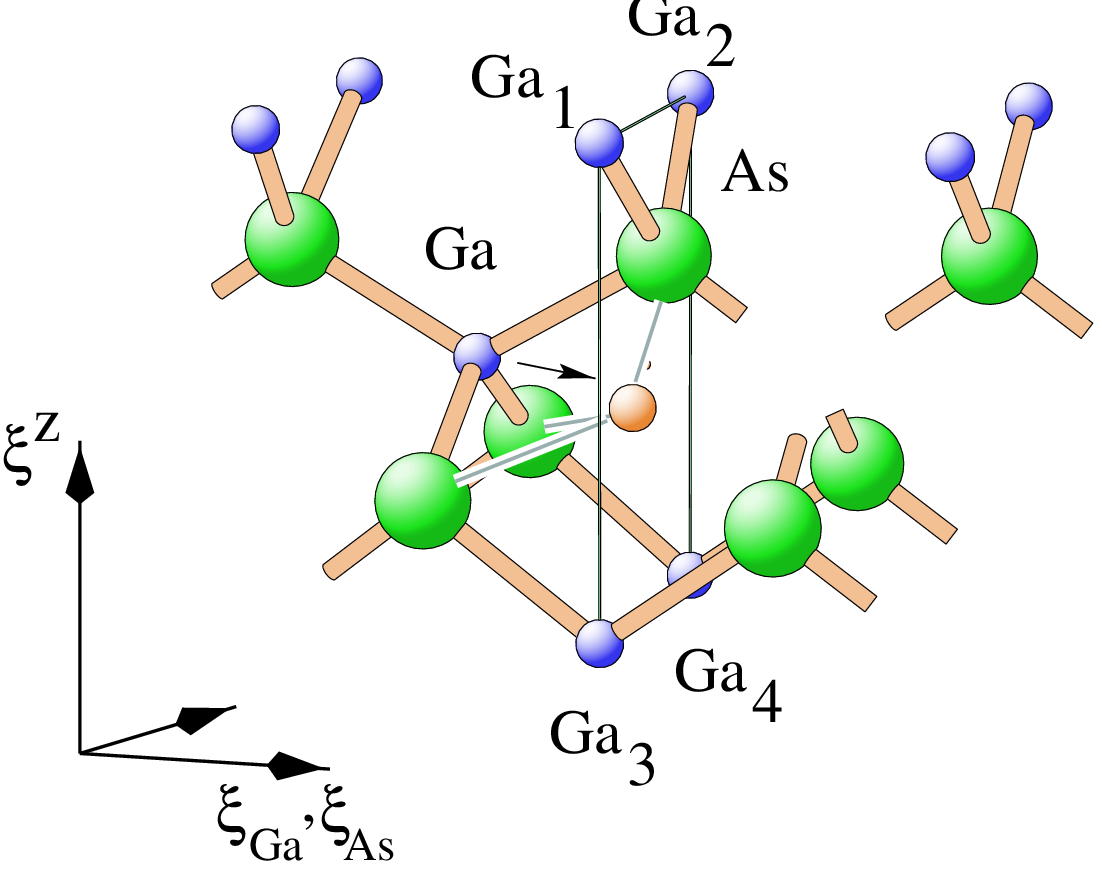,height=6cm,angle=-90}}
  \rput{0}(11.5,5){\epsfig{figure=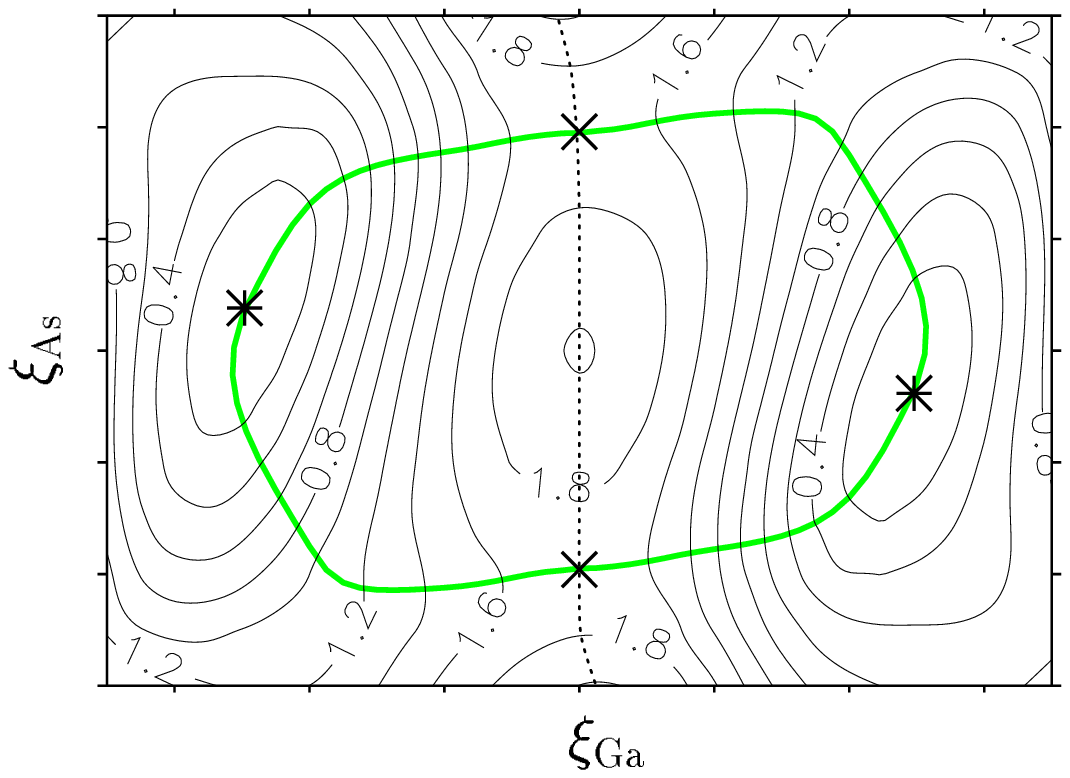,height=6cm}}
  \endpspicture
\vspace*{4cm}
\caption{\hspace*{15cm}}
\label{2nn_hop}
\end{figure}

Fig.~\ref{2nn_hop}:\\ Second nearest neighbour hop: (a) geometry. The four
atoms labelled ${\rm Ga}_1$,\ldots,${\rm Ga}_4$ form the gate through which the
hopping gallium atom has to pass. The arrow indicates the direction of the hop
and the point of return of the gallium atom found in the simulations is also
shown. (b) potential energy surface. The coordinates $\xi_{\rm Ga}$ and
$\xi_{\rm As}$ are as depicted in (a) (c.f. text). The crosses mark the saddle
points and the stars indicate the minima. The dividing surface is indicated by
the dashed line.\\[1cm] 

\begin{figure}
\psfig{file=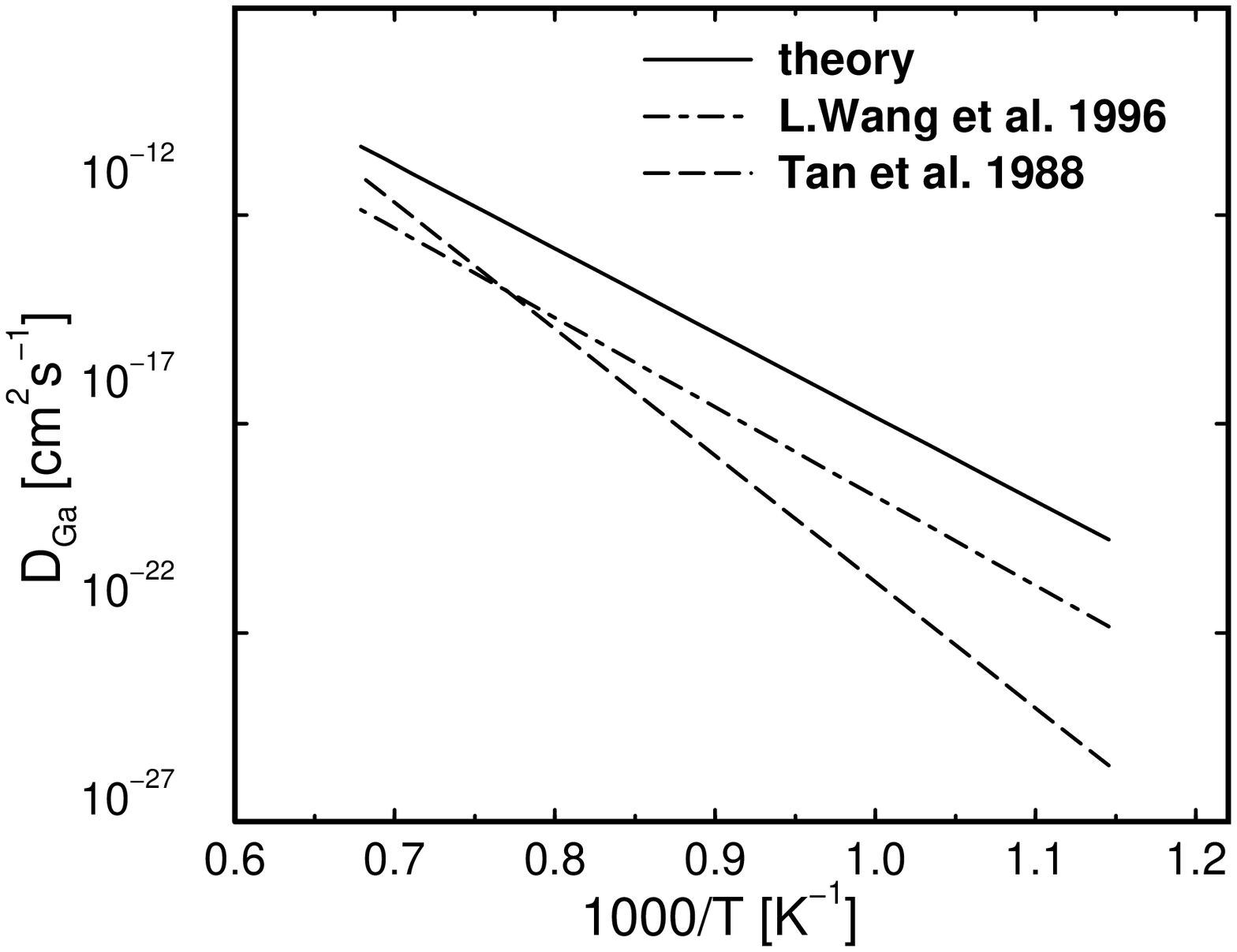,width=12.5cm}
\vspace*{4cm}
\caption{\hspace*{15cm}}
\label{diff}
\end{figure}

Fig.~\ref{diff}:\\ Gallium self-diffusion constant vs. inverse temperature.
Comparison of the theoretical result (solid) with the fit by Tan
{\em et~al.}~\cite{tan:88} to interdiffusion data (dashed) and the fit by Wang
{\em et~al.}~\cite{wang:96} to self-diffusion data (dashed-dotted).

\clearpage
\clearpage
\begin{figure}
\epsfig{file=eps.Barrier.react2.fermi,width=12cm}
\vspace*{4cm}
\caption{\hspace*{15cm}}
\label{nn_hop_fermi}
\end{figure} 
\clearpage
\clearpage
\end{document}